# Evidence for Modified Newtonian Dynamics from Cavendish-type gravitational constant experiments

Norbert Klein, Imperial College London, Department of Materials

South Kensington Campus, London SW7 2AZ, United Kingdom

*Abstract*:   Recent experimental results for the gravitational constant $G$ from Cavendish-type experiments were analysed in the framework of Modified Newtonian Dynamics (MOND).  MOND corrections were applied to the equation of motion of a pendulum, under the assumption that the magnitude of the horizontal time dependent gravitational acceleration determines the amount of MOND corrections. The large vertical component of the local gravitational field of the earth is fully compensated by the alignment of the torsion pendulum in accordance with Newton's third law and therefore not considered for MOND corrections. From the analysis of the MOND corrected equation of motion of a realistic torsion pendulum with mixed gravitational and electromagnetic restoring torque simple rules for meaningful MOND corrections of measured $G$ values determined by different operational modes of Cavendish type $G$ experiments were derived. Based on this analysis the reported discrepancies for $G$ determined by "static deflection" and "electrostatic servo" methods of the "BIPM" experiment by Quinn et al. and between time-of-swing and angular acceleration feedback methods for the "HUST" experiment by Li et al. could be fully resolved by MOND corrections using one common MOND interpolation function, determined by a one parameter fit.  The MOND corrected "BIPM" and "HUST" results, along with other "single method" results from $G$ experiments by Gundlach and Merkovitz, Schlamminger et al. and Newman et al. lead to an average $G$ value of $6.67422 \cdot 10^{-11}$ $m^3 kg^{-1}$ $s^{-2}$ with a standard deviation of 12.5 ppm only. The applied MOND correction procedure and the fitted interpolation function employed for the $G$ experiments were found to be consistent with the most viable MOND fits to galaxy rotation curves.

### 1.  Brief introduction to the MOND phenomenology

Gravity is one of the biggest challenges of fundamental physics. Although Einstein's theory of General Relativity has successfully passed any experimental test so far, culminating in the recent direct observation of gravitational waves [1], the dynamics of galaxies remains an open issue. Whereas the mainstream explanation for galaxy rotation curves is still cold dark matter (CDM) [2] - a glue which holds galaxies together without violating Newton's law of gravity - it has emerged recently that all galaxy rotation curves follow a universal law [3].  This universal relation between the observed acceleration – determined from galaxy rotation curves – and the calculated acceleration due to baryonic – i.e. visible – matter according to Newton's law is controlled by one fundamental acceleration parameter $a_0$, its numerical value is about $10^{-10} m/s^2$. CDM models so far have failed to provide a consistent explanation for this universal relation. Numerous experiments designed to detect potential dark matter particles have produced only null results to date [4]. Keeping in mind the relevance of dark matter for our current understanding of the universe, these difficulties represent a quite substantial crisis of fundamental physics, but at the same time offer a great challenge for new ideas and discoveries in the future.

As an alternative to the dark matter paradigm, modifications of Newton's law have been suggested to explain galaxy rotation curves - as a viable alternative to CDM. The most noticeable approaches are summarized under Modified Newtonian Dynamics (MOND), introduced by Milgrom in 1983 [5,6]. According to the MOND paradigm, which is purely phenomenological, the gravitational acceleration field of an isolated point mass deviates from the Newtonian acceleration $a_N$ according to

$$a_{MOND} = a_N F(|a_N|/a_0) \quad \text{with } F(|a_N|/a_0 \to \infty) = 1 \quad \text{and } F(|a_N|/a_0 \to 0) = \sqrt{a_0/|a_N|} \qquad \text{Eq. 1}$$

with $a_0 = 1.2 \cdot 10^{-10}$ m/s$^2$ denoting the fundamental MOND acceleration parameter [7]. The so-called MOND interpolation function $F$ depends on the ratio of the magnitude of the Newtonian acceleration $a_N$ to $a_0$. The constraints for the choice of $F$ are the Newtonian limit ($|a_N| >> a_0$) and the so-called "deep MOND limit" ($|a_N| << a_0$): In the deep MOND limit the acceleration field of a point mass $M$ is given by $a_{MOND} = (GMa_0)^{1/2}/r$ ($G$ = gravitational constant), which describes the observed "flat" (radius independent) galaxy rotation curves and the baryonic Tully Fisher relation [8]. Flat rotation curves are not consistent with Newton's law – without the assumption of vast amounts of dark matter within a well-matched halo around each galaxy. The $M^{1/2}$ dependence of $a_{MOND}$ in the deep MOND limit illustrates that MOND is a nonlinear theory, which has implications for the dynamical behaviour of moving masses in a gravitational field.

Except its limits, the choice of the interpolation function $F$ is not defined by any known physical law and can be determined by fits to experimental data. According to the MOND nonrelativistic field theory based on a modified Poisson equation [7] the functions dubbed "MOND simple"

$$F_{\text{MONDsimple}}\left(|a_N|/a_0\right) = \left[\frac{1}{2} + \sqrt{\frac{1}{4} + \frac{a_0}{|a_N|}}\right] \qquad \text{Eq. 2}$$

and "MOND standard"

$$F_{\text{MONDstandard}}\left(|a_N|/a_0\right) = \sqrt{\frac{1}{2} + \frac{1}{2}\sqrt{1 + \left(\frac{2a_0}{|a_N|}\right)^2}} \qquad \text{Eq. 3}$$

represent solutions of this field equation for a single point mass and have been used successfully to fit galaxy rotation curves. For clarity, it is worth to note that within the formalism set out by the modified Poisson equation the interpolation function is defined as $\mu(x)$ with $x = a_0/|a|$ rather than $x = a_0/|a_N|$. Based on this slightly different definition Equation 2 and 3 correspond to the well-known expressions $\mu_{simple}(x) = 1/(1+x)$ and $\mu_{standard}(x) = 1/(1+x^2)^{1/2}$, respectively.

Following $x = a_0/|a_N|$ recently McGaugh suggested

$$F_{\text{McGaugh}}\left(|a_N|/a_0\right) = \frac{1}{1 - \exp\left(-\sqrt{|a_N|/a_0}\right)} \qquad \text{Eq. 4}$$

as best choice to fit the plethora of galaxy rotation curves [3,4]. However, as pointed out in [10] the different functions look quite similar for the typical range of galactic acceleration magnitudes between $10^{-12}$ and $10^{-10}$ m/s² – given the large errors of experimental data (see also Fig. 7). Typical Cavendish type $G$ experiments operate in the range of $10^{-8}$ to $10^{-7}$ m/s², therefore any possible extrapolation of galaxy rotation data via MOND crucially depends on the choice of $F$. As a possible choice for a good fit to galaxy rotation curves, but allows controlling the smoothness of the transition from the deep MOND to the Newtonian limit by one parameter

$$F_{\text{Klein}}\left(|a_N|/a_0\right) = \left[1 + \left(\frac{a_0}{|a_N|}\right)^{\beta}\right]^{-\frac{1}{2\beta}} \qquad \text{Eq. 5}$$

was suggested recently with $\beta$ to be used as fit parameter [9]. As discussed in section 4 (see Fig. 7), $\beta$ values between about 0.8 and 2 cover the "smoothness" range between "MOND simple" and "MOND standard" and allow for reasonable fits to galaxy rotation curves. It is important to note that no particular choice of the MOND interpolation function is outstanding with respect to any known physical explanation of MOND effects.

The key difference between the suggested interpretations of the MOND paradigm is determined by the exact meaning of $|a_N|$ in the argument of $F$, in case of scenarios where more than one point mass is present: According to the MOND version formulated by a modified Poisson equation $|a_N|$ is the magnitude of the total gravitational field [7]. This excludes the observability of MOND effects on earth and even within our solar system. Moreover, it leads to the so-called "external field effect", which is important for MOND dynamics of satellite galaxies [11]. However, as pointed out recently MOND nonlinear field equations leads to unphysical solutions for the two-body problem [12]. Moreover, a relativistic generalization of MOND field theories has not been successful to date [13].

As a viable alternative to non-relativistic nonlinear MOND field theories Milgrom suggested in 1994 that MOND effects may result from a modification of the intertial mass of a test particle at low values of the acceleration magnitude [14], and several variants of modified inertia have been discussed [15]. Attempts have been made recently to describe modified intertia theories by a local Lagrangian [16]. Since MOND inertia interpretations do not necessarily lead to any external field effect (although some toy models exist where external fields play a role [15]) , the possibility of testing MOND in terrestrial laboratories has been considered previously: Ignatiev suggested that MOND effects may lead to spontaneous acceleration around the equinox date and suggested that experimental verification could be feasible [17]. Das and Patitsat suggested to test modified MOND effects by experiments which provide a local inertial frame of reference system, such as free fall laboratory experiments similar to the ones being used for tests of the weak equivalence principle [18]: In their contribution the authors claim that torsion balance experiments are not suited for this purpose because the centripetal acceleration resulting from the earth translates to the apparatus being used. However, since a torsion balance is based on a pendulum which is aligned to the direction of the vector sum of the gravitational acceleration of the earth and the centripetal force, I argue that the conditions are very close to a local inertial frame of reference, if the pendulum is sufficiently decoupled from seismic noise and Brownian motion, i.e. if the pendulum body is kept sufficiently quiet.

## 2. MOND corrections for a torsion pendulum with mixed gravitational and electromagnetic restoring torque

Fig. 1 shows the schematics of a modern torsion pendulum experiment being used to determine *G*. According to the "modified inertia" interpretation of MOND the acceleration magnitude $|a_N|$ to be employed for MOND corrections describes the component of the gravitational field which leads to an accelerated motion of a test mass. In case of an ideal torsion pendulum the plane perpendicular to the pendulum fibre represents an approximate two-dimensional inertial frame of reference - for the limit of infinite pendulum length. Gravitational fields of moon and sun, the centrifugal acceleration due to the rotation of the earth and local gravitational sources like buildings contribute to the local gravitational field vector ($g_{local}$ in Fig. 1) which defines the pendulum alignment.

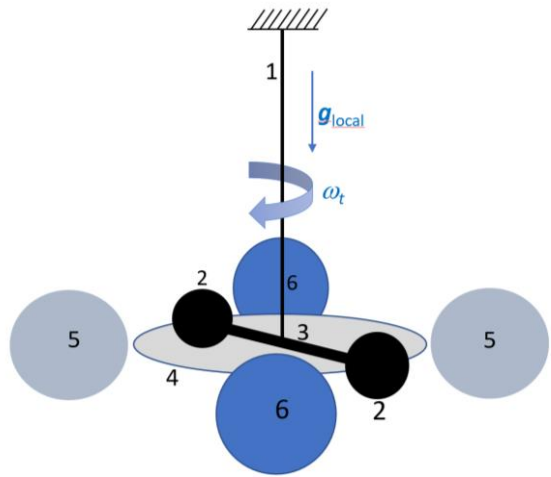

Fig.1: Schematics of a Cavendish type *G* experiment. A suspended torsion fibre or torsion strip (1) is employed as suspension of a test mass, ideally composed of two spherical masses (2) which are connected by a massless rigid bar of length *L* (3). The plane oriented perpendicular to the local effective gravity vector $g_{local}$ (4) represents an approximate 2D inertial frame of reference (4). Two source masses are arranged with their centre-of-mass within this plane (4) and can be moved between near (5) and far position (6), aiming to generate a torque to the torsion wire, which is measured via angular deflection of the pendulum in order to determine *G*. In case of the AFF method (see text) the pendulum suspension rotates at a constant angular velocity $\omega_t$ around its axis.

Torsion pendulum experiments are incredible sensitive and have been used to test violations of Newton's second law at acceleration magnitudes as low as $10^{-13}$ m/s$^2$: According to results reported by Gundlach et al. [19] deviations from Newton's second law can be excluded for acceleration magnitudes as low as $10^{-14}$ m/s$^2$, but only for electromagnetic forces causing the acceleration of the pendulum body - in this case the restoring torque from the torsion fibre which originates from the elastic properties of the fibre material.

Based on this experimental constraint the only credible way to implement MOND corrections into the differential equation for a pendulum (torsion or linear alike) is given by Eq. 6

$$\frac{d^2\vec{r}_{//}(t)}{dt^2} = (\vec{g}_{//}(t) + \vec{a}_{centri}(t)) \cdot F\left(\frac{\left|\vec{g}_{//}(t) + \vec{a}_{centri}(t)\right|}{a_0}\right) + \vec{a}_{em,//}(t) \qquad \text{Eq. 6}$$

- as approximation for an idealized pendulum body composed of two point masses *m* connected by a massless bar of length *L* according Fig. 1. In Eq. 6 $\vec{r}_{//}(t)$ describes the horizontal position of one of the two pendulum masses, exposed to a time dependent horizontal gravitational field $\vec{g}_{//}(t)$ and to a horizontal electromagnetically generated acceleration $\vec{a}_{em,//}$. The centrifugal acceleration $\left|\vec{a}_{centri}(t)\right| = \frac{L}{2} \cdot \omega_t^2(t)$ arises in case of an enforced rotation of the pendulum suspension with an angular frequency $\omega_t(t)$ (see Fig. 1), as being used in case of the "Angular Acceleration Feedback" (AAF) method (see section 3). According to General Relativity a local gravitational field is equivalent to an enforced accelerated motion, therefore the vector sum $\vec{g}_{//}(t) + \vec{a}_{centri}(t)$ determines the effective horizontal gravitational field, its magnitude determines the amount of MOND correction.

Eq. 6 represents the "MOND equivalence" to Newton's equation of motion as normally used to describe the equilibrium between the gravitational force to be measured and the counterforce provided by the *G*-apparatus, the latter may contain electromagnetic forces (for example the restoring torque of the torsion fibre), but also gravitational forces (the gravitational part of the restoring torque and a possible "in-plane" accelerated motion of the pendulum body). Although to date there exists no fully fledged version of MOND which would enable the analysis of a complex arrangement like a (rotating) torsion pendulum with mixed gravitational and electromagnetic restoring torque from one action, Eq. 6 represents a consistent empirical approach, which takes into account existing experimental constraints about nil-observation of MOND effects in case of small electromagnetic forces, as reported by Gundlach et al. [19]. In fact, Eq. 6 is consistent with "MOND" fits to galaxy rotation curves: here $\vec{a}_{em} = \vec{a}_{centri} = 0$, hence Eq. 7 is identical to the MOND modified inertia paradigm (the index // for "horizontal" is omitted) set out by a modification of Newton's second law at small acceleration magnitudes.

$$\frac{m}{F(\left|\vec{g}(t)/a_0\right|)}\frac{d^2\vec{r}(t)}{dt^2} = m\vec{g}(t) \qquad \text{Eq. 7}$$

As an important side remark, Eq. 6 and 7 do not take into account the gravitational field of the pendulum mass *m*, in other words, the source masses are assumed to be large in comparison to *m*. This approximation is valid for most of the Cavendish type experiments. As pointed out recently, for the two-body problem the most viable approach is to pursue MOND corrections for the relative acceleration [9,12]. In fact, MOND inertia models for multi body problems required for accurate modelling of galaxy clusters or dwarf galaxies are still lacking. The RAMSES code being used for multibody problems within MOND [20] relies on a quasi-linearized version of the modified Poisson equation (QUMOND) formulated by Milgrom in 2010 [21]. Although this version has technical advantages over the Modified Poisson equation, it does not include the inertia interpretation of MOND.

In order to describe the motion of the pendulum for small torsional acceleration magnitudes as a one-dimensional problem for the motion of a point mass, I chose a Cartesian coordinate system with z-axis aligned with the pendulum axis and with *x*-axis oriented parallel to tangential direction at the azimuthal equilibrium position $\varphi_0$ of one of the two pendulum masses. Small torsional oscillations of the pendulum can be described by $x(t) = L/2 \cdot \varphi(t)$ resulting in a one-dimensional differential equation for $x(t)$ according to Eq. 8:

$$\ddot{x}(t) = \left[ g_{ext,x}(t) - \omega_0^2(1-\chi)x(t) \right] \cdot F\left( \left| \sqrt{\left( g_{ext,x}(t) - \omega_0^2(1-\chi)x(t) \right)^2 + \left( \frac{L}{2}\omega_t^2 \right)^2} \right| / a_0 \right) +$$

$$a_{em,x}(t) - \omega_0^2 \chi x(t) - \frac{\omega_0}{Q}\dot{x}(t)$$

Eq. 8

The first term in brackets describes the *x*-component of the gravitational field, which is the sum of the horizontal gravitational field $g_{ext,x}(t)$ generated by the enforced motion of the source masses during a run of a *G*-experiment, the second term describes the gravitational portion of the restoring force of the pendulum, which is proportional to $x(t)$: Like for a real Cavendish experiment it is assumed here that the restoring torque of the pendulum is a mixture of a gravitational torque (due to the gravitational field of the earth or due to the gravitational potential of the source masses, see section 3). The parameter $0 \le \chi \le 1$ describes the relative amount of electromagnetic $\kappa_{em}$ restoring torque of the torsion fibre,

$$\chi = \frac{\kappa_{em}}{\kappa_g + \kappa_{em}}$$

Eq. 9

$\omega_0$ is the angular resonance frequency of the torsion pendulum for the case of zero damping ($Q \to \infty$) and $\kappa_g$ the gravitational restoring torque. According to the definition of $\chi$ by Eq. 9 and presumptions set out by Eq. 6 the electromagnetic portion of the restoring force $-\omega_0^2 \chi x(t)$ does not require any MOND correction. The same holds true for the damping term which is of electromagnetic nature. For the case that the pendulum suspension is rotating with a constant angular velocity $\omega_t$ during the operation of a *G*-experiment, the magnitude of the sum of the gravitational and centrifugal acceleration determines the amount of MOND correction according to Eq. 6, the quadratic sum in the argument of *F* in Eq. 8 results from the fact that the two components are oriented perpendicular to each other. As discussed in section 3, MOND corrections of AAF type Cavendish *G* results can be pursued using the peak-to-peak value of the gravitational and acceleration magnitudes which occur during a complete cycle of a given experiment.

### 3. MOND corrections for four different operational modes of Cavendish "big G" experiments

| No | $G/(10^{-11}$ $m^3kg^{-1}s^{-2})$ | error [ppm] | $a/a_0$ | Method | $\chi = \dfrac{\kappa_{em}}{\kappa_g + \kappa_{em}}$ | Reference |
|---|---|---|---|---|---|---|
| 1 | 6.674184 | 11.64 | $\approx 202$ | Cavendish ToS | $\approx 0.99$ torsion wire | Li et al. 2018 [22] |
| 2 | 6.674484 | 11.61 | $1275 = (1261^2+202^2)^{1/2}$ | Cavendish AAF | $\approx 0.99$ torsion wire | Li et al. 2018 [22] |
| 3 | 6.67433 | 19 | 9733 | Cavendish ToS | $\approx 0.99999$ torsion wire | Newman et al. 2014 [23] |
| 4 | 6.67586 | 54 | 396 | Cavendish SD | $\approx 0.03$ torsion strip | Quinn et al. 2014 [24] |
| 5 | 6.67515 | 61 | 396 | Cavendish ESS | $\approx 0.03$ torsion strip | Quinn et al. 2014 [24] |
| 6 | 6.674252 | 24.4 | $g_0$ not included: 6533 ; $g_0$ included: $8\cdot10^{10}$ | beam balance | n.a. | Schlamminger et al. 2006 [25] |
| 7 | 6.674215 | 13.8 | $5139 = (5136^2+176^2)^{1/2}$ | Cavendish AAF | $\approx 0.99$ torsion wire | Gundlach, Merkowitz 2000 [26] |

Table 1: $G$ – results from Cavendish experiments published over the last ten years. No 1,2,4,5 represent results where more than one operational mode of a given experiment were compared. Earlier results by the groups of Li et al. (see [27,28]) and Quinn et al. ([29,30]) are not listed. For comparison, the Cavendish result by Gundlach and Merkowitz from 2000 is listed (no 7), because it represents the original AAF method which was adapted by Li et al. (no 1). As an example of a non-Cavendish experiment, which relies on a commercial beam balance and does not employ any purpose-developed detection scheme, the 2006 result by Schlamminger et al. (no 6) is included in the table. As discussed in detail in section 3, the listed acceleration magnitudes (in units of the MOND acceleration $a_0=1.2\cdot10^{-10}$ m/s$^2$ ) for each experiment which were employed for the MOND corrections was calculated from data given in the references. In case of the AAF method the acceleration is presented as $a = (a_c^2+g^2)^{1/2}$ which $a_c$ denoting the radial centripetal acceleration due to the rotation of the pendulum turntable and $g$ the azimuthal acceleration due to the gravitational field of the source masses. In contrast to Cavendish experiments, where the gravitational field of the earth $g_0$ is perpendicular to the gravitational field by the source masses, both fields are parallel in case of the beam balance experiment (no 6), leading to two possible values for the acceleration magnitude, depending on whether $g_0$ is considered or not considered for MOND corrections. In order to label the applied operational modes of the Cavendish experiments the abbreviations ToS are used for "time of swing", AAF for "angular acceleration feedback", SD for "static deflection" and ESS for "electrostatic servo" (see text and references therein), respectively. The parameter $\chi$ represents the fraction of the electromagnetic restoring torque of the torsion pendulum according to Eq. 9. The $G$ values listed in this table - along with MOND corrected values are graphically presented in section 4 (Fig. 6).

Aiming to tackle the uncertainties due to the elastic properties of the torsion fibre, Quinn et al. have employed a torsion strip, which enables the use of a larger test mass of ca. 6.6 kg, rather than ca. 100 g for most of the torsion fibre-based experiments. Any twist of a torsion strip leads to a small lift of the test mass in the gravitational field of the earth, which results in a restoring torque which is largely determined by the magnitude of this lift. According to Eq. 8 such a pendulum can be described by a $\chi$ value close to zero. In other words, the torsion strip pendulum by Quinn et al. is a gravitational pendulum, with the big advantages of a low "gravitational spring constant" and weak excitation by seismic noise - in comparison to a linear gravitational pendulum. A detailed description of the different operational methods listed in Table 1 (except No 1 and 2) is given in the review by Rothleitner and Schlamminger [31] and in the original publications. In the following the basic principles of these methods and their specific MOND correction are discussed based on Eq. 8. In order to work out the required MOND correction for the static deflection and the time-of-swing methods as a function of the parameter $\chi$, numerical solutions of the dynamical pendulum behaviour according to Eq. 8 will be discussed.

The most simple case with regards to MOND corrections is the "Electrostatic Servo" (ESS) method: here the pendulum body does not move at all because the gravitational torque generated by moving source masses between "far" and "near" position is compensated by an electromagnetic torque at any time of a given experimental cycle. The electromagnetic torque is generated by a voltage applied to the capacitors, which is determined via a feedback loop from the measured pendulum deflection. For this "static case" Eq. 8 can be drastically simplified:

$$0 = g_{ext,x}(t) \cdot F\left(\left|g_{ext,x}(t)\right|/a_0\right) + a_{em,x}(t) \Rightarrow \left|\frac{a_{em,x}(t)}{g_{ext,x}(t)}\right| = F\left(\left|g_{ext,x}(t)\right|/a_0\right) \qquad \text{Eq. 10}$$

As a consequence of Eq. 10, *G* values determined by the ESS method are expected to be "MOND enhanced" by the value of the MOND interpolation function *F* at the maximum acceleration magnitude of the pendulum between "near" and "far" position of the source masses. Since Eq. 10 is independent of $\chi$, this result is independent of the nature of the restoring torque of the torsion fibre or torsion strip. Consequently, MOND corrections for EES – based *G* results can be calculated according to Eq. 11

$$\frac{G_{MOND,ESS}}{G_{ESS}} = F\left(\left|g_{ext,pp}\right|/a_0\right) \qquad \text{Eq. 11}$$

with $g_{ext,pp}$ denoting the "peak-to-peak" magnitude of the azimuthal differential gravitational field between near and far position, which is listed in table 1.

A similar methodology of MOND corrections needs to be applied for *G* values determined by the "angular acceleration feedback" (AAF) method, originally developed by Gundlach and Merkowitz [26] and later compared with the "time-of-swing" (ToS) method by Li et al. [22]. Here the measured pendulum deflection during the cyclic motion of the source masses between "near" and "far" position (in this case a rotation of the source masses around the pendulum axis with constant angular velocity) is compensated by a feedback loop. In contrast to the EES experiment the pendulum suspension is located on a turntable, and the feedback from the measured pendulum deflection drives an accelerated azimuthal motion of pendulum turntable aiming to compensate the gravity-induced deflection of the torsion pendulum from its equilibrium position (details of the AAF method are explained in [22,26]). Eq. 8 describes the experiment in a rotating (with the angular frequency $\omega_t$ of the pendulum turntable) non-inertial frame of reference. Like for the case of the EES method, the restoring force terms are zero and the measured azimuthal acceleration $\ddot{x}(t)$ needs to be MOND corrected:

$$\ddot{x}(t) = g_{ext,x}(t) \cdot F\left(\left|\sqrt{g_{ext,x}{}^2(t) + \left(\frac{L}{2}\omega_t{}^2\right)^2}\right| / a_0\right) = \frac{L}{2}\dot{\omega}(t) \cdot F\left(\left|\sqrt{g_{ext,x}{}^2(t) + \left(\frac{L}{2}\omega_t{}^2\right)^2}\right| / a_0\right) \quad \text{Eq. 12}$$

In Eq. 12 $\dot{\omega}(t)$ is the measured angular acceleration of the turntable which exhibits a sinusoidal time dependence due to the circular motion of the test masses and the pendulum – for the case of the point mass approximation according to Fig. 1.

Since the selected frame of reference is a non-inertial frame of reference, each test mass feels a centrifugal acceleration, and the modulus of the vector sum of azimuthal and centrifugal acceleration determines the relevant magnitude for MOND corrections (see experimental results in table 1). The time dependent centrifugal acceleration due to the accelerated azimuthal motion and the minuscule centrifugal acceleration due to the earth rotation are negligible.

Similar to the case of ESS, the measured peak-to-peak angular acceleration of the turntable $\dot{\omega}_{PP}$ due to the gravitational field by a full rotation of the source masses with respect to the pendulum turntable determines the magnitude of MOND corrections.

$$\frac{G_{MOND,AAF}}{G_{AAF}} = F\left(\sqrt{\left(g_{ext,pp}\right)^2 + \left(a_{centri}\right)^2} / a_0\right) = F\left(\sqrt{\left(\frac{L}{2}\dot{\omega}_{PP}\right)^2 + \left(\frac{L}{2}\omega_t{}^2\right)^2} / a_0\right) \quad \text{Eq.13}$$

In contrast to the ESS and AAF methods, the static deflection (SD) and time-of-swing (ToS) modes of operation utilize oscillations of the pendulum, i.e. the restoring force terms in Eq. 8 cannot be neglected. Therefore one may expect that in this case MOND corrections depend on the parameter $\chi$. In order to analyse the dynamical pendulum behaviour, numerical calculations of the MOND corrected pendulum motion $x(t)$ (without pendulum rotation) were performed using the iteration ($i=0$ to $N_{max}$)

$$\ddot{x}(t_i) = \left[g_{ext,x}(t_i) - \omega_0{}^2(1-\chi)x(t_i)\right] \cdot F\left(\left|g_{ext,x}(t_i) - \omega_0{}^2(1-\chi)x(t_i)\right| / a_0\right) - \omega_0{}^2\chi x(t_i) - \frac{\omega_0}{Q}\dot{x}(t_i)$$

$$\dot{x}(t_{i+1}) = \dot{x}(t_i) + \ddot{x}(t_i)(t_{i+1} - t_i)$$

$$x(t_{i+1}) = x(t_i) + \dot{x}(t_i)(t_{i+1} - t_i)$$

Eq.14

for a given set of start parameters $x(t_0 = 0) = 0$ and $\dot{x}(t_0 = 0) = 0$ using a MATLAB script. In order to reduce numerical errors to an acceptable level, in particular for the determination of small changes of the pendulum resonance frequency due to MOND effects, $N_{max} \approx 1$ million was found to be sufficient for the calculation of $x(t)$ over 5-10 pendulum periods. For $g_{ext}(t)$ a ramp was chosen starting with $g_{ext}(t=0) = 0$ followed by a linear increase of $g(t)$ towards a constant value $g_0$ for $t \geq T_{ramp}$. It turned out that the new equilibrium position taken by the pendulum at $t > T_{ramp}$, which is equal to $\omega_0{}^{-2}g_0$ in the Newtonian limit, is independent of $T_{ramp}$ - in case of MOND corrections being included. Moreover, a nonlinear $g_{ext}(t)$ did not change the new equilibrium position. This is a non-trivial statement because of the nonlinear character of MOND.

Fig. 2 shows selected examples of simulations. In order to visualize MOND effects, a low acceleration value of $g_0 = a_0 = 1.2 \cdot 10^{-10}$ m/s$^2$ was chosen. The chosen MOND interpolation function according to Eq. 5 with $\beta = 1.25$ results in a MOND enhancement of the acceleration at $a_N = a_0$ by a factor 1.321.

The amplitudes $x$ are presented in units of $\omega_0^{-2} g_0$, i.e. $x = 1$ represents the Newtonian case. It is important to note that the general picture is independent of the choice of the interpolation function.

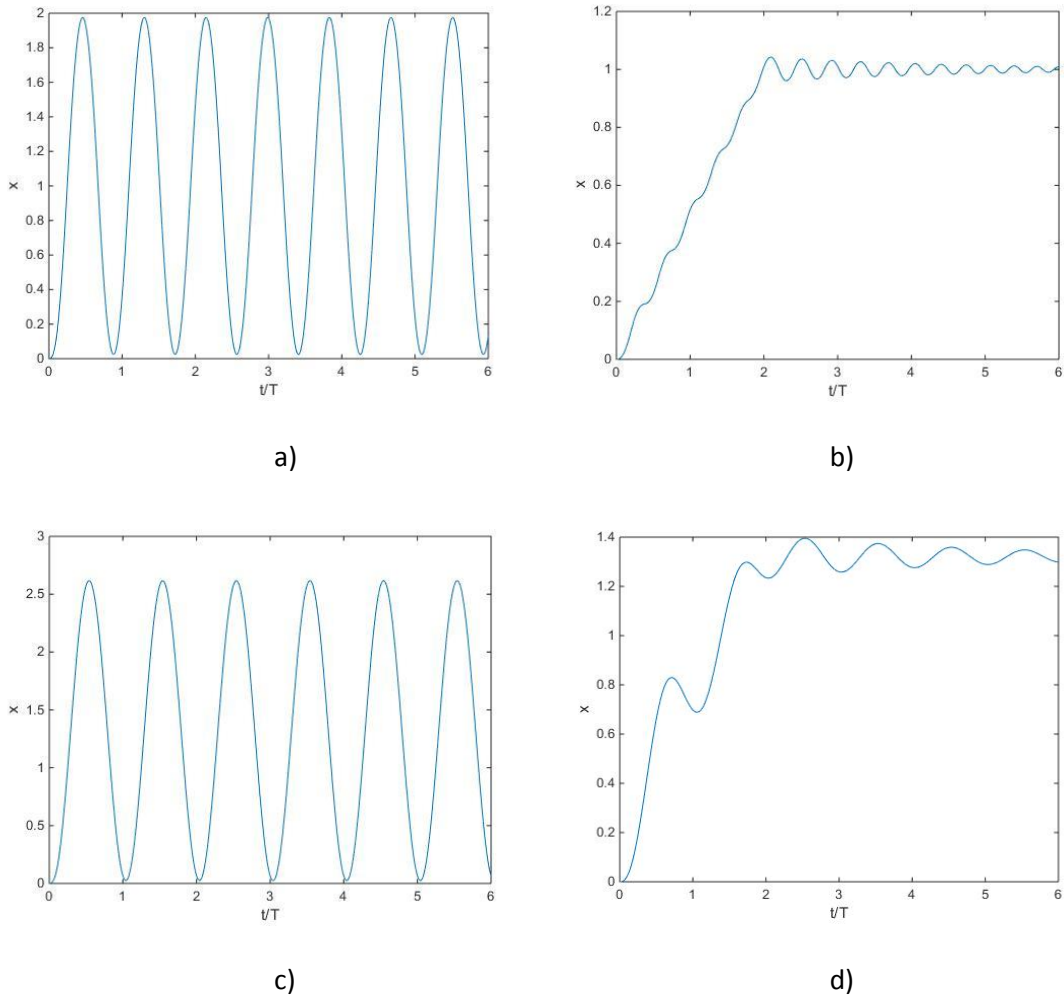

a)                                                                 b)

c)                                                                 d)

**Fig. 2:** Results of numerical calculations based on Eq. 14 at $g_0 = a_0$. Displacements $x$ in units of $a_0/\omega_0^2$. Static enhancement with respect to Newton: $g_{MOND}/g_{Newton} = 1.321$.
a) Gravitational restoring force, short ramp ($T_{ramp} = 0.2 \cdot T$), no damping ($Q = 10^5$)
b) Gravitational restoring force, long ramp ($T_{ramp} = 2 \cdot T$), strong damping ($Q = 3$)
c) Electromagnetic restoring force, short ramp ($T_{ramp} = 0.2 \cdot T$), no damping ($Q = 10^5$)
c) Electromagnetic restoring force, long ramp ($T_{ramp} = 2 \cdot T$), strong damping ($Q = 3$)

Fig. 2a shows the results for the case of a pure gravitational restoring force ($\chi = 0$) for nearly zero damping ($Q = 10^5$) and a short ramp ($T_{ramp} = 0.2 \cdot T$) with $T$ denoting the pendulum period $T = 2\pi/\omega_0$ in the Newtonian limit. The time is presented in units of $T$. The simulation reveals a time dependence of $x(t)$ visually indistinguishable from sinusoidal, but with a frequency which is noticeable larger than in the Newtonian limit. Like in the Newtonian limit, the pendulum still oscillates around $x = 1$. Fig. 2b shows the result of the simulation for a gravitational pendulum with strong damping ($Q = 3$) and $T_{ramp}$ equal to two pendulum periods. The results reveal a further increase of the pendulum frequency in comparison to Fig. 2a, but the Newtonian value for the equilibrium value of the oscillation for $t > T_{ramp}$ is retained. The frequency of the oscillation is directly related to its amplitude.

Figs. 2c and d show the simulated response of the pendulum for the case of an electromagnetic restoring force ($\chi = 1$) for zero (c) and strong damping (d). In this case the pendulum resonance frequency retains the Newtonian value, but the equilibrium position for $t > T_{ramp}$ is increased by an amount $F_{MOND}(|g_0|/a_0)$, which is 1.321 for the given example. As a consequence, in this case the same amount of MOND correction arises as in case of the ESS method (see Eq. 11).

The important non-trivial case is a gravitational pendulum ($\chi =0$): MOND does not affect the change of the pendulum equilibrium position. The physical reason is that the balance between external gravitational force and restoring force, the sum of both is MOND corrected (see Eq. 8). However, there is one caveat which leads to a very specific MOND correction of the static deflection $G$-results for a gravitational pendulum: In order to convert the measured deflection angle into a torque, from which $G$ is determined, the restoring torque coefficient $\kappa \approx \kappa_g$ needs to be determined experimentally. Usually $\kappa$ is determined from a measurement of the pendulum frequency $\omega_0$ [24]. Since the pendulum amplitude determines the MOND-related frequency increase $\Delta\omega_0$ of the pendulum, the corresponding relative increase of $\kappa$ is given by

$$\frac{\Delta\kappa}{\kappa} = 2 \cdot \frac{\Delta\omega_0}{\omega_0}.$$

Eq. 15

The "factor 2" in Eq. 16 results from $\omega_0 \propto \kappa^{1/2}$. As result, the MOND correction for the SD method of a gravitational pendulum is

$$\frac{G_{MOND,SD}}{G_{SD}} = 2 \cdot \frac{\Delta\omega_{PP}\left(F\left(\left|g_{ext,pp}\right|/a_0\right)\right)}{\omega_0}$$

Eq. 16

with $\Delta\omega_{PP}\left(F\left(\left|g_{ext,pp}\right|/a_0\right)\right)/\omega_0$ representing the MOND-induced change of the pendulum resonance frequency for the given choice of the MOND interpolation function $F$ and $g_{ext,pp}$, which is determined by the peak-to-peak gravitational torque during one cycle of the experiment (source masses moved from near to far position or via versa). The last statement is only exact for the case that the pendulum does not oscillate prior to the gravitational excitation, a larger pendulum amplitude may diminish MOND effects. The relative frequency change $\Delta\omega_{PP}\left(F\left(\left|g_{ext,pp}\right|/a_0\right)\right)/\omega_0$ cannot be calculated analytically, but by a numerical calculation according to Fig. 3a based on the iteration given by Equation 14 for $g_0 = g_{ext,pp}$ and the selected interpolation function, with $g_{ext,pp}$ taken from table 1. As an example for a particular choice of the MOND interpolation function (Eq. 5 with $\beta = 1.26$) Fig. 3 compares the expected MOND $G$-corrections (expressed as $(G_{MOND}-G)/G$ ) for the two cases as a function of the magnitude of $g_{exp,pp}$ according to Eq. 11 ($\chi = 1$: SD, ESS, $\chi = 0$: ESS) with Eq. 16 ($\chi = 0$: SD). Each point in Fig. 3 (squares) was determined by numerical calculations. The results demonstrate that there is a different amount of MOND correction for $G$ determined by SD in comparison to ESS – for the case of a gravitational pendulum ($\chi = 0$). The crossover between the two curves at a relative acceleration amplitude of about 0.3 has no simple intuitive explanation. For the acceleration range of $G$ experiments listed in table 1 Eq. 16 leads to a larger amount of MOND correction for SD experiments.

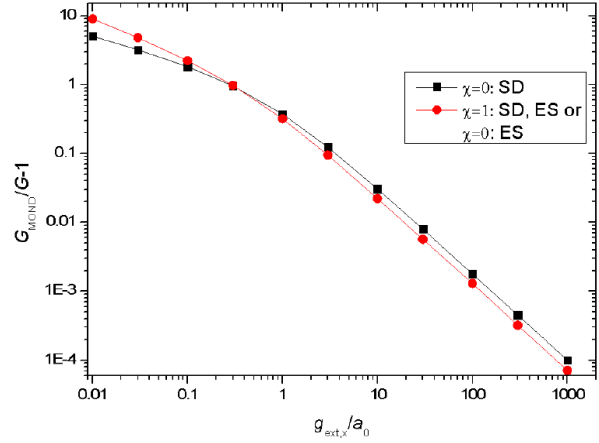

Fig. 3: Calculated expected MOND correction according to Eq. 11 (circles) and Eq.16 (squares) as a function of the magnitude of the horizontal gravitational field for the MOND interpolation function according to Eq. 5 with $\beta$ = 1.26. For the values calculated by Eq. 16 the iteration according to Eq. 14 was employed for the case of zero damping ($Q \rightarrow \infty$).

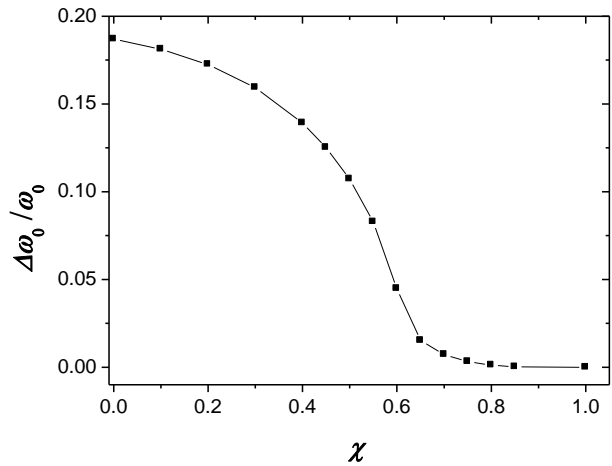

Fig. 4: Calculated relative frequency change of a pendulum due to MOND as a function of the relative fraction of the electromagnetic portion $\chi$ of the pendulum restoring force coefficient for a pendulum acceleration amplitude of $a_0$.

The case of a mixed gravitational/electromagnetic restoring force is of particular interest for the so-called "time-of-swing" (ToS), which has gained popularity in recent years. As explained in detail in [22,31], this method relies on measurements of small changes of the resonance frequency of the pendulum for two different positions of the source masses: In the "near" position the additional gravitational force between the source masses and the pendulum body generates a small gravitational component of the restoring torque coefficient, which has to be added to the electromagnetic torque coefficient of a fibre-based torsion pendulum. It is of special interest to evaluate the MOND-induced frequency changes as a function of $\chi$. The results of the simulation, again for $g_0=a_0$ and the same MOND interpolation function used in Figs. 2 and 3, are displayed in Fig. 4.

The predicted frequency change shows a strongly non-linear variation with $\chi$, and nearly disappears for $\chi > 0.8$. As an important consequence of this surprising result, one cannot expect that the time-of-swing method is sensitive to MOND corrections, as long as the gravitational component of the restoring torque is a few percent of the total one only, which is the case for all ToS experiment listed in table 1.

$$\frac{G_{MOND,ToS}}{G_{ToS}} = 1$$ 
Eq. 17

This result shows that the ToS method is an ideal method for accurate *G* measurements – unaffected by MOND corrections – for the case of a fibre torsion pendulum. Until now, no time-of-swing experiment employing a torsion strip has been reported. In contrast to a torsion fibre, these experiments would be sensitive to MOND effects.

### 4. Results of MOND fits to *G* experiments

MOND corrections of the *G*-data presented in table 1 are based on the "flexible" MOND interpolation function according to Eq. 5. The only reason that this particular choice has been used is the fact the transition from the deep MOND limit to the Newtonian limit can be smoothly varied by a one parameter fit using $\beta$ as fit parameter. The numerical value of the MOND acceleration parameter $a_0$ is known from fits to the Tully Fisher relation [8], therefore a one parameter fit is sufficient. Apart from the method-dependent choice of the MOND correction procedure according to Eqs. 11, 12, 16 and 17, which was discussed in detail in the previous section, the key parameter which defines the amount of MOND correction for a given choice of $\beta$ is the acceleration amplitude in units of $a_0$, as listed in table 1. Before discussing the results of the MOND corrections displayed in Fig. 5, it is described below how the listed values of $a/a_0$ were estimated from the information given in the original publications:

**No 1 and 2 Li et al.:** Recently Li et al. reported a discrepancy for *G* measured by two operation methods of a conventional torsion experiment which employs a torsion wire rather than the torsion strip [22]. This result breaks previous records in terms of the quoted measurement error. In spite of the fact that the experiment has been operated and improved over many year, the averaged *G* values determined by the angular acceleration feedback (AAF) method is significantly higher than the one determined by the time-of-swing (ToS) method. In fact, *G* determined by AAF is 45 ppm higher than *G* determined by ToS, although the quoted measurement error is 11.6 ppm for both.

The AAF method allows a direct measurement of the angular acceleration. According to [22], the peak-to-peak amplitude is about 924 nrad/s². The pendulum body is of rectangular shape, the width (which determines the strength of the measured angular deflection) is $b$ = 91 mm. In order to pursue MOND corrections, I was considering an equivalent point mass pendulum body (see Fig. 1) composed of two point masses on a radius $R$ such that the moment of inertia for a torsional rotation $I_{pm} \approx m \cdot R^2$ is equal to that of the rectangular pendulum body $I_{rec} \approx 1/12 \cdot m \cdot b^2$ of the same mass. The resulting Newtonian linear acceleration magnitude comes out to be $g = 202 \cdot a_0$. Due to the rotation of the pendulum turntable at a quoted angular velocity $\omega_t$ = 2.44 mrad/s the corresponding centrifugal acceleration $a_{centri} = \omega_t^2 \cdot b/12^{1/2} = 1261 \cdot a_0$. Since $a_{N,source\ masses}$ and $a_{N,centri}$ are perpendicular to each other, the estimated total magnitude of acceleration $|a| = (g^2 + a_{centri}^2)^{1/2} = 1277 \cdot a_0$ (see table 1). Therefore, owing to the rotating turntable MOND effects are largely supressed, but still not negligible because of the high accuracy of the experiment.

In case of the ToS method the period of the pendulum is decreased by 1.7 s in the "near" position with respect to the "far" position of the source masses [22]. Given the average pendulum period of $T \approx 430$ s and the fact that the restoring force of the pendulum is 100% electromagnetic in the far "position", $\chi$ comes out to be 0.9961 according to Eq. 9. Calculations of the pendulum frequency with this value of $\chi$ do not show any measurable deviations from Newton. Therefore, the $G$ value measured by ToS does not require any MOND correction.

**No 4 and 5 Quinn et al. :** In this experiment a copper beryllium torsion strip rather than a torsion fibre has been used. This leads to the already mentioned nearly 100% gravitational character of the restoring torque. In fact, the values reported in [24] for the gravitational and electromagnetic components are $\kappa_g = 2.18 \cdot 10^{-4}$ Nm/rad and $\kappa_{em} = 7.5 \cdot 10^{-6}$ Nm/rad, respectively, leading to $\chi = 0.033$ according to Eq. 9. Simulations of the pendulum period based on the iteration given by Eq. 14 have shown that for these small values of $\chi$ the results are not very sensitive to the exact numerical value, therefor even $\chi = 0$ does not change the results in any noticeable way. The exact choice of the pendulum quality factor at the given values between 10,000 and 100,000 has no influence on the results within the numerical error margins.

The pendulum body used in the experiment is composed of four cylindrically shaped Cu-Te cylinders of $m_t$=1.2 kg arranged in a "quadrupole" configuration on top of an aluminium plate. The radius of the circle which intersects with the axis of each cylinder is $R_1$ = 120 mm. The source mass is formed by an array of four cylinders of ca. $m_s$=11 kg each, which are arranged on a turntable. At an angle $\varphi_0$ = 18.9° between field and source masses the torque created by the gravitational force between source and field masses is at its maximum. In order to conduct $G$ measurements, the source mass turntable is moved periodically between - 18.9 ° and + 18.9°, leading to torque of 3.1489·10⁻⁸ Nm applied to the pendulum [16]. With the given moment of inertia of the pendulum body of $I$ = 7.9598·10⁻²kgm² the angular acceleration of the pendulum body is $d^2\varphi/dt^2 = torque / I$ = 3.9561·10⁻⁷s⁻². Given the radius $R_{tm}$ = 120 mm of the test mass carousel (test masses are the major contribution to the moment of inertia mass of the pendulum body), the relevant linear acceleration magnitude for this experiment is $g \approx R_{tm} \cdot d^2\varphi/dt^2$ = 4.75·10⁻⁸s⁻² = 396·$a_0$ (see table 1).

**No 6 Schlamminger et al. :** A quite unique experiment where the weight difference of two pendulum bodies is determined in the presence of two seven ton mercury source masses [25]. The weight is determined by a modified commercial beam balance, such that systematic detector errors are more unlikely. Therefore, this experiment may be considered as a gold standard and is therefore included in the analysis. The estimation of the acceleration magnitudes is straightforward - based on the quoted gravitational forces and the pendulum mass - resulting in an acceleration magnitude of 6533·$a_0$ (see table 1). In the framework of the MOND interpretation being used for the analysis of G experiment it remains questionable whether the $G$ values from this experiment needs any MOND correction at all.

**No 7 Gundlach and Merkowitz:** Similar to Li et al, a rectangular shaped pendulum body of width $L$ = 76 mm is employed, the measured peak-to-peak angular acceleration value is 960 nrad/s² [26], but the angular velocity of the pendulum turntable $\omega_t$ = 5.3 mrad/s is significantly higher than for Li et al., yielding $|a| = (g^2 + a_{centr}^2)^{1/2} = (176^2 + 5136^2)^{1/2} \cdot a_0 = 5138 \cdot a_0$ (see table 1).

The acceleration magnitude of the experiment by Newman et al. (No 3, [23]) was determined in a similar way, but the exact value is not relevant for the analysis because ToS—based *G* values do not require any MOND correction.

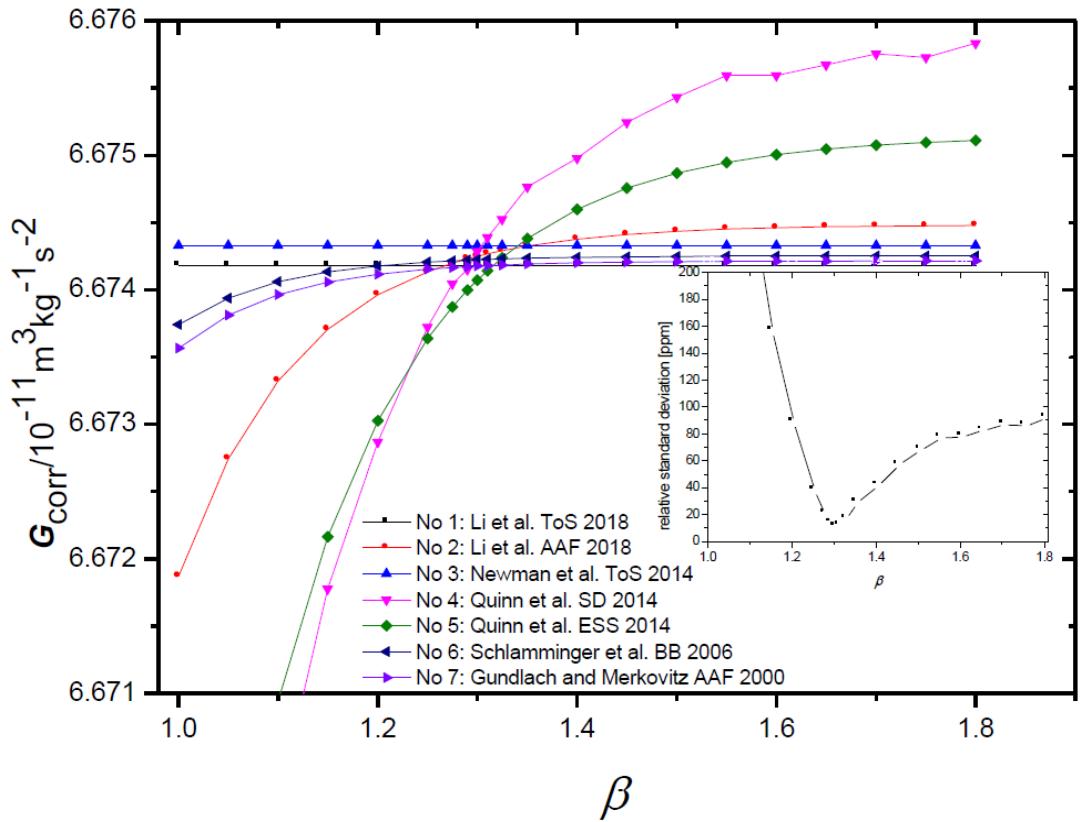

Fig. 5: MOND corrected *G*-values determined from the experimental data shown in table 1 based on the flexible MOND interpolation (Eq. 5) function for different value of the numerical fit parameter β. ToS results do not require MOND corrections (Eq. 17). MOND corrections for ESS results were calculated by Eq. 11, for AAF by Eq.13 and for SD by Eq. 16. For the calculations the acceleration magnitudes listed in table 1 were used. The beam balance results were corrected under the assumption that the gravitational background field $g_0$ of the earth is not considered for MOND corrections (upper entry in table 1). The scattering of data points for corrected data by Quinn et al SD (No 4) is due to the numerical error of about $5·10^{-6}$ for the determination of the MOND corrected pendulum resonance frequency based on the iteration given by Eq. 14. The insert shows the relative standard deviation of the corrected *G*-values with a sharp minimum at β = 1.30, leading to an average *G* value of $6.67422·10^{-11}$ m³ kg⁻¹s⁻² with a relative standard deviation of 12.5 ppm.

Fig. 5 shows the results of the analysis. The MOND corrected *G*-values determined from the experimental data listed in table 1 based on the flexible MOND interpolation (Eq. 5) function for different value of the numerical fit parameter β is displayed. ToS results do not require any MOND correction (Eq. 17) and appear as horizontal lines. MOND corrections for ESS results were calculated by Eq. 11, for AAF by Eq.13 and for SD by Eq. 16. For the calculations the acceleration magnitudes listed in table 1 were used. The beam balance results were corrected under the assumption that the gravitational background field $g_0$ of the earth is not considered for MOND corrections (upper entry in table 1). The scattering of data points for the corrected data by Quinn et al. SD (No 4) is due to the numerical error of about $5·10^{-6}$ for the determination of the MOND corrected pendulum resonance frequency based on the iteration given by Eq. 14. The insert shows the relative standard deviation of the corrected $G$-values with a sharp minimum at $\beta$ = 1.30, leading to an average $G$ value of $6.67421 \cdot 10^{-11}$ m$^3$ kg$^{-1}$s$^{-2}$ with a relative standard deviation of 12.5 ppm. For large values of β the MOND corrections become negligible and the $G$ values correspond to the uncorrected ones. Here the standard deviation is ca. 85 ppm, therefore the MOND corrections lead to a significant reduction of the data scattering of big $G$. Fig. 6 shows a comparison of experimental and MOND corrected $G$ values including the quoted error bars for the best fit with $\beta$ = 1.30. Since the quoted error bars include estimated systematic errors, which are not rigorously determined by one common methodology for the different experiments, I did not include any weighting according to the error bar size for the determination of the average $G$ and standard deviation. According to Fig. 6 MOND corrections explain the large $G$ values reported by Quinn et al., and the reported discrepancy between the two different modes of operation. Although the reported difference between the two methods reported Li et al. is significantly smaller than for Quinn et al. on an absolute scale, it is important to note that MOND corrections with one common interpolation function explain the reported discrepancies within the error bars of the experiments. The small MOND corrections to be applied for the reported $G$ values by Gundlach and Merkovitz and by Schlammrner et al. are within their error bars. However, with regards to the consistency of the analysis it is very important that these results do not contradict with the MOND analysis. The fact that the average value of $G$ is so close to the "gold standard" experiment by Schlammrner et al. provides additional confidence that the MOND analysis is a large step forward towards a possible explanation of the "big $G$" conundrum.

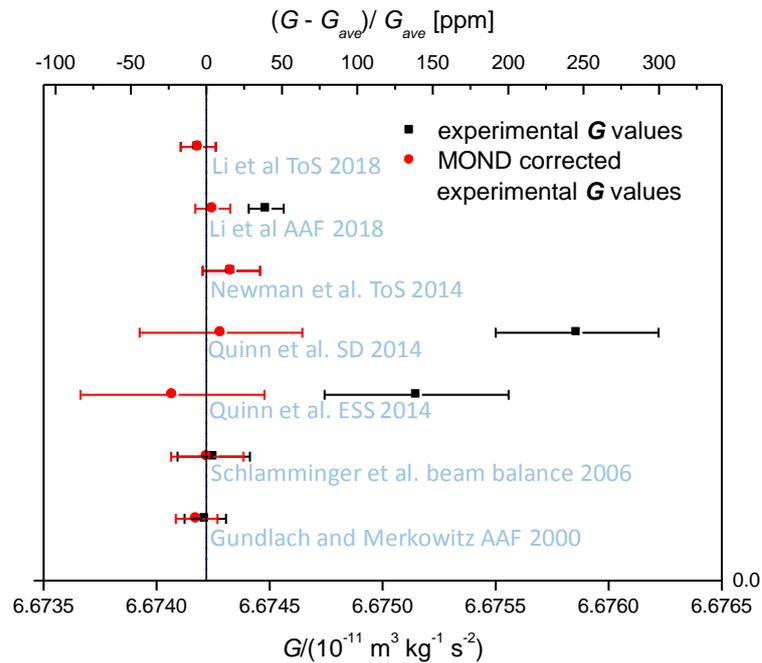

Fig. 6. Comparison of measured $G$ values with MOND corrected values according to Eq. 5 for $\beta$ = 1.30. The vertical line represents the average $G$ values after MOND correction of $6.67422 \cdot 10^{-11}$ m$^3$ kg$^{-1}$s$^{-2}$ with a relative standard deviation of 12.5 ppm. The error bars for both the original and MOND corrected $G$ values represent the errors quoted by the authors of the original papers. The quoted average $G$ value and its standard deviation does not include any weighting of data points according to the given error bars.

Fig. 7 shows the MOND correction $\Delta a/a_N$ as a function of the magnitude of the Newtonian acceleration $a_N$ (in units of $a_0$) for the SD and ESS mode of the BIPM experiment, and for the average AAF results reported by Li et al. The error bars represent the experimental errors. For a direct comparison with recent astrophysical data, the purple dots represent individually resolved measurements along the rotation curves of nearly 100 spiral galaxies. The original data in [32] are presented as ratio of the squares of the measured and calculated orbital velocities - the latter from the Newtonian gravitational acceleration by the baryonic (= visible stars and interstellar gas) mass of the galaxy. The full lines represent the MOND interpolation function according to Eq. 5 for different values of the fit parameter $\beta$, including fits to the measured data from the two $G$ experiments. Fig. 7 will be further discussed in section 5.

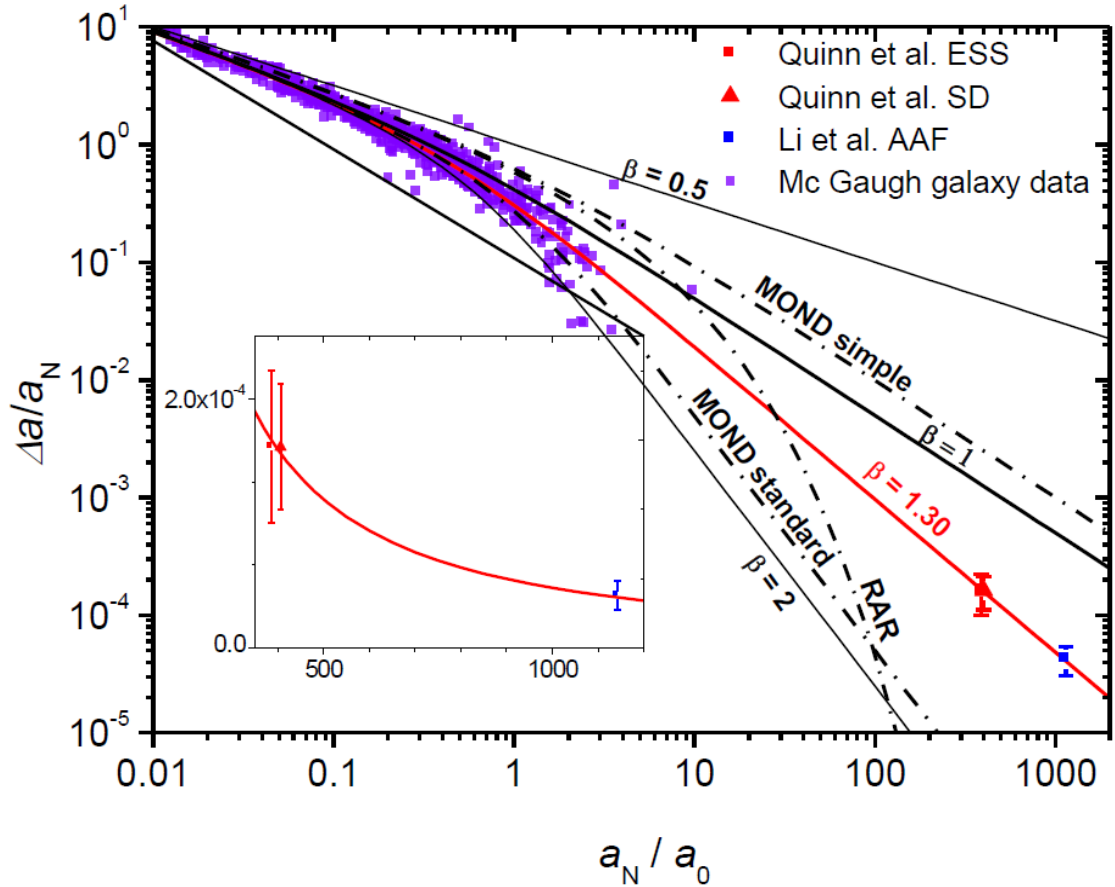

Fig. 7: Relative deviation from Newtonian gravitational acceleration as a function of the magnitude of the Newtonian acceleration in units of $a_0$ for the two operational modes of the experiment by Quinn et al. [24] and the AFA mode of the experiment by Li et al. [22]. In order the visualize the BIPM results determined by the two methods, the electric servo and static deflection results are displayed with their identical $a_n/a_0$ value moved to the left and right, respectively. The insert shows a magnification of the diagram around the data points from the $G$ experiments. The purple data points represent data extracted from galaxy rotation curves according to [24]. The MOND interpolation functions $F_{Klein}$ (Eq. 5) for several choices of the parameter $\beta$ (full lines, including $\beta = 1.30$ obtained from fits to the $G$ experiments) are compared with two common MOND interpolation functions (Eqs. 2 and 3) and Mc Gaugh's universal radial acceleration relation (RAR, Eq. 4).

### 5. Discussion

The analysis of a pendulum at small acceleration amplitudes within a MOND inertia scenario presented in this contribution shows that different operational modes of a given $G$ experiment can lead to different results for the measured $G$ values. This unique quality of the MOND paradigm distinguishes MOND from other suggested modifications of Newton's law, such as short range corrections by a Yukawa term to be added to the Newtonian gravitational potential. For the latter tight constraints on the strength and the relevant length scales were revealed from a variety of experiments - as part of the "Eöt-Wash" campaign [33]. It is important to emphasize that the observed differences in $G$ from one experiment employing different modes of operation are not consistent with any Yukawa type modified gravity, because the values of the source masses and distance range of the two versions are almost identical. In contrast, the observed differences could be explained by the dynamical behaviour of a given Cavendish $G$-apparatus, which is determined by the unique nonlinear nature of MOND. As an example, any Yukawa-type modification of the gravitational force would not require any correction of the restoring force of the pendulum in the experiment by Quinn et al, because this force is just a component of the gravitational field of the earth, its magnitude is well inside the excluded range of Yukawa-type modifications of Newton's law. Therefore, it is exclusive to the MOND paradigm that the presented analysis of recent $G$ experiments provides a consistent explanation of the observed discrepancies of $G$ results determined from 7 different experiments employing 5 different methods. This remarkable result provides experimental evidence for Modified Newtonian Dynamics from terrestrial experiments for the first time.

The comparison of the observed discrepancies with galaxy rotation curves and MOND interpolation functions displayed in Fig. 7, although appealing in terms of the perfect matching of the fitted interpolation functions with the average of the rotation curve results, should be taken with a pinch of salt: As mentioned before, the choice of a particular MOND interpolation function is not motivated by any known physical mechanism. Therefore, the observed consistency of the fitted interpolation function with the galaxy data does suggest - but not prove - that the observed deviations from Newton have the same physical origin than the rotation curves of galaxies. Fig. 7 should be taken as working hypothesis, aiming to refine and design experiments which operate closer to the acceleration range of galaxies. In fact, the results published by Gundlach et al. in 2007 suggest that this is possible [19]. A straightforward way to provide further evidence based on existing experiments is to run the AAF experiment by Li et al. at lower speed of the pendulum turntable, which should lead to higher values of $G$, if MOND effects are taken into account. In case of the gravitational torsion pendulum used by Quinn et al. precise measurements of the pendulum period as a function of amplitude towards – as close as possible towards $a_0 / \omega_b^2$ - may prove or disprove the MOND hypothesis. However, this is not as trivial as it sounds, because the influence of parasitic effects like temperature drift, noise and environmental gravitational gradients may cause increasing measurement errors.

The observed deviations from Newton are not in direct conflict with the experimental limits for possible deviations from Newton within our solar system: The acceleration magnitude of $a_N/a_0 \approx 400$ for the BIPM experiment is equal to the gravitational acceleration of the sun at a distance of 380 AU (astronomical units), which is about ten times the distance between sun and Pluto. However, the extrapolation of these results via the fitted MOND interpolation function to the distance of Pluto and Saturn leads to a MOND correction of $5.5 \cdot 10^{-7}$ and $1.3 \cdot 10^{-8}$, respectively. This is just about equal to the upper $2\sigma$ exclusion boundary for anomalous radial acceleration in case of Saturn and Uranus, and not in conflict with other planets of our solar system (see table 2 in [34]).

With regards to the assumed interpretation of MOND, the analysis shows strong evidence that MOND effects can indeed be observed in the presence of a gravitational field which is much larger than Milgrom's acceleration parameter $a_0$, as long as the dynamical degrees of freedom are confined to a 2D plane with normal vector strictly parallel oriented to the local external gravitational field (see Fig. 1). In this context it is important to note that fits of galaxy rotation curves using the RAR (Eq. 4) are based on the MOND inertia interpretation: as discussed in [3] these fits were pursued by employing the ordinary Poisson equation for the calculation of the radial baryonic Newtonian acceleration $g_{b,Newton}$ from the observed matter distribution, subsequently using $F|g_{b,Newton}/a_0|$ as MOND correction. This procedure is equivalent to Eq. 7 and therefore consistent with the analysis of $G$ experiments presented here. In fact, from a recent comprehensive study of fits to individual SPARC galaxy rotation curves by McGaugh et al. a preference of the modified inertia over the modified Poisson version of MOND was reported [35]. Therefore, the common statement that "MOND effects cannot be observed on earth", which is a direct consequence of the MOND modified Poisson nonlinear field theory [7], may be wrong. The presented analysis of $G$ experiments by MOND is fully consistent with the most viable MOND fits to galaxy rotation curves.

Finally, the MOND analysis of $G$ experiments provides evidence that electromagnetic forces (in this case the restoring torque of a torsion wire) are not subject to MOND corrections, which confirms the findings from previous work [19] that a general MOND modification of the inertial mass can be ruled out. This is not in any conflict with the success of "MOND inertia" to explain galaxy rotation curves, because there are no electromagnetic fields contributing to the dynamical behaviour of galaxies. In the context of General Relativity, the exclusion of electromagnetically driven accelerations may be interpreted as a hint that tiny deviations from flat 2D spacetime due to in-plane gravitational fields may control the amount of MOND corrections. This possible scenario is consistent with modified inertia in case that no electromagnetic forces contribute to the dynamical behaviour of a system, but it suggests that electromagnetic forces are excluded from MOND corrections.

**Conclusions**

The observed discrepancies between values of the gravitational constant determined by different operational modes of recent Cavendish-type experiments were found to be consistent with Modified Newtonian Dynamics and MOND fits to galaxy rotation curves. The fact that MOND explains different $G$ results determined by different operational modes of a given experiment differentiates MOND from other possible modifications of Newton's law, such as Yukawa terms – as predicted by supersymmetry theories. Beyond this exclusive evidence for MOND, the re-analysis of most credible recent $G$ experiments by MOND has lead to a significant reduction of data scattering of "big-$G$" determined by the most credible recent $G$-experiments and suggests that the "real" value of $G$ is $6.67422 \cdot 10^{-11}$ m$^3$kg$^{-1}$s$^{-2}$ – within a standard deviation for the corrected experimental data of only 12.5 ppm. This finding could mark a consolidation of "big $G$" metrology within the experimental error margins of individual experiments and may solve the conundrum about the least accurate know constant of nature. Future experiments with improved sensitivity at smaller acceleration amplitudes should be pursued to support these initial findings and to fill the gap between the acceleration magnitude of galaxy rotation curves and current terrestrial *G* experiments. The paper describes the methodology of data analysis according to the MOND paradigm and puts restrictions on possible physical interpretations of the MOND phenomenology. The indicated opportunity for the direct experimental verification of Milgrom's Modified Newtonian Dynamics at the acceleration scale of galaxies in earth-bound laboratories provides an amazing perspective: If MOND effects at the scale of $a_0$ should be verified by "small-scale" terrestrial experiments, ΛCDM cosmology would require a significant modification because the explanation of galaxy dynamics by cold dark matter – already increasingly challenged by the predictive power of MOND with regards to galaxy dynamics – would finally be falsified. Such a possible finding would impose a gigantic challenge on our current understanding of the universe.

### 6. Acknowledgements

I would like to express my sincere thanks to Hinrich Meyer, now emeritus Professor at Bergische Universitaet Wuppertal (Germany), where I graduated from. His group is still running a linear double pendulum experiment at DESY in Germany, which I originally designed in 1987 as part of my PhD thesis. Hinrich Meyer has been looking for MOND effects in his double pendulum experiment and published data which exclude MOND, but without taking into account the dynamic effects of MOND [36]. This result has inspired me to take a closer look at Cavendish type *G* experiments. I would like to thank Clive Speake, Professor at University of Birmingham, who is running the BIPM experiment, which is currently located at NIST. Clive gave me the opportunity to visit this outstanding experiment in his laboratories in Birmingham during 2016, and without this great experience and his explanations I would not have been able to pursue this analysis in a meaningful way. Finally, I like to thank Dr Stephan Schlamminger from NIST for helpful discussions about my analysis and for referring to an older version of my manuscript in his recent publication about measurement errors in "big *G*" experiments [37].

### 7. References

[1] B. P. Abbott *et al.* (LIGO Scientific Collaboration and Virgo Collaboration), Phys. Rev. Lett. **119**, 161101 ( 2017)

[2] A. Baudis, European Review **26**, 70 (2017)

[3] S. McGaugh, F. Lelli, J.M. Schombert, Phys. Rev. Lett. **117**, 201101 (2016)

[4] P. Li et al., Astronomy and Astrophysics **615**, A3 (2018)

[5] M. Milgrom, Astrophys. J. **270**, 365 (1983)

[6] B. Famaey, S. McGaugh, Living Rev. Relativity **15**, 10 (2012)

[7] J.D. Bekenstein, M. Milgrom, Astrophys. J. **286**, 7 (1984)

[8] S. McGaugh, The Astronomical Journal **143**, 40 (2012)

[9] N. Klein, arXiv:**1504.07622v4** (2016)

[10] D.C. Rodrigues et al, Nature Astronomy **2**, 668 (2018)

[11] E. Garaldi et al., Phys. Rev. Lett. **120**, 261301 (2018)

[12] J. Klacka, arXiv:**1904.04074** (2019)

[13] see for example B. Famey and S. Mc Gaugh, Living Reviews in Relativity **15**, 10 (2012), and references therein.

[14] M. Milgrom, Annals of Physics **129**, 384 (1994)

[15] M. Milgrom, Acta Physica Polonica B **42**, 2175 (2011)

[16] R. Costa, G. Franzmann, J.P. Pereira, arXiv: **1904.07321** (2019)

[17] A. Ignatiev, Phys. Rev. Lett. **98**, 101101 (2007) and Canadian Journal of Physics **93**, 166 (2015)

[18] S. Das, S.N. Patitsat, Phys. Rev. D **87**, 107101 (2013)

[19] J.H. Gundlach et al., Phys. Rev. Lett. **98**, 105801 (2007)

[20] F. Lueghausen, B. Famaey, P. Kroupa, Canadian Journal of Physics **93**, DOI: 10.1139/cjp-2014-0168 (2014)

[21] M. Milgrom, MINRAS **403**, 886 (2010)

[22] Q. Li et al., Nature **560**, 582 (2018)

[23] R.Newman et al., Phil. Trans. R. Soc. A **372**, 20140025 (2014)

[24] T. Quinn et al., Trans . R. Soc. A **372**, 20140032 (2014)

[25] S. Schlamminger et al, Phys. Rev. D **74**, 082001 (2006)

[26] J.H. Gundlach, S.M. Merkowitz, Phys. Rev. Lett. **85**, 2869 (2000)

[27] L.C. Tu et al., Phys. Rev. D **82**, 02201 (2010)

[28] Q. Li et al., Phil. Trans. R. Soc. A **372**, 20140141 (2014)

[29] T.J. Quinn et al., Phys. Rev. Lett. **87**, 111101 (2001)

[30] T. Quinn et al., Phys. Rev. Lett. **111**, 101102 (2013) and Phys. Rev. Lett. 113, 039901 (Erratum) (2014)

[31] C. Rothleitner, S. Schlamminger, Review of Scientific Instruments **88**, 111101 (2017)

[32] B. Famaey, S. Mc Gaugh, Living Rev. Relativity **15**, 10 (2012), Fig 10, bottom panel, data on http://astroweb.case.edu/ssm/data

[33] T.A. Wagner, S.Schlamminger, J.H. Gundlach, E.G. Adelberger, Classical and Quantum Gravity **29**, 18 (2012)

[34] M. Sereno, Ph. Jetzer, Mon. Not. R. Astron. Soc. **371**, 626 (2006)

[35] P. Li et al., Astronomy & Astrophysics **615**, A3 (2018)

[36] H. Meyer et al., Gen. Relativ. Gravit. **44**, 2537 (2012)

[37] C. Markatas et al., Metrologia **56**, 0549001 (2019)